\documentstyle[epsf]{l-aa}
\begin{document}

%%%%%%

\def\h2{H\,{\sc ii}}
\def\MC{Magellanic Cloud}
\def\sm{$M_{\odot}$}
\def\sl{$L_{\odot}$}
\def\sr{$R_{\odot}$}
\def\T{$T_{eff}$}
\def\L{$(L/L_{\odot})$}
\def\r{$(R/R_{\odot})$}
\def\kms{km s$^{-1}$}
\def\sk{Sk$-66^{\circ}41$}
\def\h27{HDE\,269227}
\def\R{R\,84}
\def\frac{$''$\hspace*{-.1cm}}
\def\min{$'$\hspace*{-.1cm}}
\def\s{star}
\def\O{Ofpe/WN9}
\def\l{$\lambda$}
\def\fc{Fe\,{\sc{iii}}}
\def\fb{Fe\,{\sc{ii}}}
\def\ha{H\,{\sc{i}}}
\def\hea{He\,{\sc{i}}}
\def\heb{He\,{{\sc ii}}}
\def\nb{N\,{\sc{ii}}}
\def\nc{N\,{\sc{iii}}}
\def\nd{N\,{\sc{iv}}}
\def\ne{N\,{\sc{v}}}
\def\ca{Ca\,{\sc{ii}}}
\def\sib{Si\,{\sc{ii}}}
\def\sic{Si\,{\sc{iii}}}
\def\sid{Si\,{\sc{iv}}}
\def\cb{C\,{\sc{ii}}}
\def\cc{C\,{\sc{iii}}}
\def\cd{C\,{\sc{iv}}}
\def\fc{Fe\,{\sc{iii}}}
\def\p{P\,Cyg}
\def\ac{Al~{\sc{iii}}\ }
\def\na{Na~{\sc{i}}\ }
\def\nf{Ne~{\sc{i}}\ }
\def\ca{Ca~{\sc{ii}}\ }
\def\soc{S~{\sc{iii}}\ }

%%%%%%%

\thesaurus{08.23.2; 08.09.2 \R; 08.05.1; 08.05.2; 08.06.3; 11.13.1}
\title{The LMC transition star \R\, and the core of the 
LH\,39 OB association\thanks{\hspace*{-.2cm}Based on observations obtained 
at the 
European Southern Observatory, La Silla, Chile.}}
\author{M.\ Heydari-Malayeri\inst{1}  \and  F.\ Courbin \inst{2,4} \and  
G. Rauw\inst{2,}\thanks{\hspace*{-.2cm}Aspirant au Fonds
National de la Recherche Scientifique (Belgium)}  \and 
O.\ Esslinger\inst{3}  \and P. Magain\inst{2,}\thanks{Ma\^\i tre de  
Recherches au Fonds
National de la Recherche Scientifique (Belgium)} } 
\offprints{M.\ Heydari-Malayeri}
\institute{{\sc demirm}, Observatoire de Paris, 61 Avenue de l'Observatoire, 
 F-75014 Paris, France \and 
Institut d'Astrophysique, Universit\'e de Li\`ege, 5, Avenue de Cointe, 
   B-4000 Li\`ege, Belgium \and 
Department of Physics and Astronomy, University of Wales, College of 
   Cardiff, CF2 3YB, Cardiff, UK 
  \and {\sc daec-ura}\,173, Observatoire de Paris, F-92195 Meudon Principal 
   cedex, France }
\date{Received date; accepted date}
\maketitle
\markboth{M.\ Heydari-Malayeri et al.: \R}{}

%\begin{center}
%{\LARGE \bf \underline{Not Final Version}}
%\end{center}

\begin{abstract} 
On the basis of sub-arcsecond imaging  obtained at 
the ESO NTT with SUSI and the ESO ADONIS adaptive optics system 
at the 3.6\,m 
telescope, we resolve and study 
the core components of the LMC OB association LH\,39. 
The central star of the association, the rare transition object 
\R, is also investigated  using CASPEC echelle spectroscopy at the ESO 
3.6\,m telescope.  
A new, powerful image restoration code that conserves the 
fluxes allows us to obtain  the magnitudes and colors of the 
components. We bring out some 30 stars in a $\sim$\,16\frac\,$\times$16\frac\, 
area centered on \R. At a resolution of \ 0\frac.19 ({\sc fwhm}),   
the closest   components to \R\, are shown to be stars 
\#21 and \#7 lying at 1\frac.1 NW and 1\frac.7 NW respectively of 
the transition star. The 
former is possibly a blue star of $V$\,=\,16.7 mag and the latter with  
its $V$\,=\,17.5 mag is the reddest star of the field, after \R. 
Star \#7 turns out to be too faint to correspond to the red M2 supergiant 
previously 
reported to contaminate the spectrum of \R.   
If the late-type spectrum is due to a line-of-sight supergiant 
with a luminosity comparable to  \R, it should lie closer than 
0\frac.12  to \R. 
The transition star shows 
spectral variability between 1982 and 1991. We also note some slight radial 
velocity variations of the Of emission lines over timescales of several 
years. Furthermore, we derive the spectral types of two of the brightest 
stars of the cluster, using long slit spectra obtained at the 
NTT telescope equipped with EMMI, and discuss the apparent absence of O 
type stars in this association.

\end{abstract}

\section{Introduction}

In spite of considerable advance achieved in the past decade on 
the formation and evolution of massive stars, 
the evolutionary sequence, between O 
and Wolf-Rayet (W-R) stars, is still far from being established. 
Of prime importance for this topic is the investigation of a 
very small family of objects, 
the so-called \O\, transition  stars, which seems to 
hold the keys for better understanding the physical 
characteristics of several massive star subclasses   
populating the upper part of the H-R diagram.  
\O\, \s s show the spectral features of both emission-line O \s s and the 
later  W-R types of the nitrogen sequence, i.e. a combination 
of high- (\heb\, and \nc) and low- (\hea\, and \nb) excitation emission 
features.  The designation  
\O\, underlines the difficulty in distinguishing between these two 
subclasses. In fact, these  \s s were given either OIafpe or 
WN9--10 classifications by Walborn (\cite{Wal82}) but were 
later revised to Ofpe/WN9 by Bohannan \& Walborn (\cite{Boh89}). 
Recently, three of the \MC\, ``slash'' \s s 
(\R, BE\,381, and HDE\,269927c) 
have been reclassified as WN9 by Crowther et al. (\cite{Cro95}).  

\parindent=1cm
Anyhow, one of the most important aspects of these    
upper H-R diagram objects  is their close link    
with the Luminous Blue Variable (LBV) 
phenomenon  (Stahl et al. \cite{Sta83}) which,   
according to current massive star evolutionary models 
(Maeder \cite{Maeder},  Langer et al. \cite{Langer}) represents a short 
stage in the evolution of O \s s initially more massive than 
60 \sm\, before the 
advent of the Wolf-Rayet phase. In fact \O\, \s s have been described as 
quiescent LBVs \, (Bohannan \& Walborn \cite{Boh89},  Crowther 
et al. \cite{Cro95} ).

\parindent=1cm
At present, only 10 massive transition  stars are known in the LMC 
(Bohannan \& Walborn \cite{Boh89}). 
This small group  should be investigated 
from every  angle in order to gain insight into 
their status. For example, the 
stellar environment of these stars and their  possible belonging   
to a  massive star  cluster deserves consideration. 
 This new approach to 
the study of the so-called \O\, \s s will be applied in this 
paper to \R\, (Feast et al. \cite{Fea}).  On the basis of  
sub-arcsecond images in the $U$, $B$, $V$, $R$  and the near infrared 
$H$ and $K$ bands, we will  try to resolve \R\, and its  
neighborhood. Furthermore, we will, for the first time, give  
accurate photometry of the unknown resolved components. 
 \R\, is of particular interest since it is the only \s\, of this 
class showing the signature of a late-type companion in its red and 
infrared spectroscopy and photometry (Allen \& Glass \cite{All76}, 
Wolf et al.\ \cite{Wol87}, McGregor et al.\ \cite{McG88}). 
Presently, we do not know whether the red companion, classified by 
Cowley \& Hutchings (\cite{Cow}) as an M2 supergiant, is physically 
related to the Ofpe/WN9 or their association is just a line-of-sight 
effect.

\parindent=1cm
\R\,   has several other designations, 
mainly: \h27\, (Henry Draper 
(Extension) catalog), Brey\,18 (Breysacher \cite{Br81}), 
Sk--69$^{\circ}$79 (Sanduleak \cite{San70}), S\,91 (Henize 
\cite{Hen56}), WS\,12 (Smith \cite{Smi68}). 
It lies in the central part of the OB association LH\,39 (Lucke \& 
Hodge \cite{LH}, star \#12) towards the southern edge of the bar.   
In this direction is also situated   
the LMC weak, filamentary H$\alpha$ nebulosity DEM\,110 (Davies et al. 
{\cite{Dav}). Feast et al. (\cite{Fea}) classified it as Pec(uliar). Later, 
it got other classifications, as follows: 
WN8 (Smith \cite{Smi68}), OIafpe (Walborn \cite{Wal77}), WN9--10 
(Walborn \cite{Wal82}), 
\O\, (Bohannan \& Walborn \cite{Boh89}), WN9 (Crowther et al. \cite{Cro95}). 
The star has been found to show significant brightness and color 
variations (Stahl et al. \cite{Sta84}) as well as moderate spectroscopic 
variability in H$\gamma$ and \hea\,\l4471  
(Stahl et al. \cite{Sta85}). 
A detailed spectral analysis of \R\, 
 was presented by Schmutz et al. (\cite{Sch91}) using 
the observations of Stahl et al. (\cite{Sta85}). On the basis of  a 
non-LTE model for a spherically expanding atmosphere, 
which was the first of the kind for an \O\, star, they derived 
the stellar parameters of \R. Hydrogen was found to be very depleted 
and it was found that \R\, had lost about half of its initial mass, and 
was probably a post-red supergiant. 
Recently Crowther et al. (\cite{Cro95}), using an independent model but 
the same set of data, have confirmed Schmutz et al.'s (\cite{Sch91}) 
findings.    

\section{Observations and data reduction}
\subsection{Sub-arcsecond imaging and deconvolution photometry}

\R\, was observed using the ESO New Technology Telescope (NTT) 
during two runs. The best images were taken  on 1991 December 26 using 
the SUperb Seeing Imager (SUSI) which
functions with an active optics system (see ESO Web site for more
information). The observing conditions were excellent with the seeing 
varying between 0\frac.50 and 0\frac.80\, ({\sc fwhm}). 
The detector was a Tektronix 
CCD (\#25) with 1024$^{2}$ pixels of 24 $\mu$m. The filters used 
(their ESO numbers, central wavelengths, bandwidths),  
the exposure times, the dates, and the pixel size on the sky are summarized 
in Table 1.

\begin{figure}
%\picplace{8 cm}
\begin{center}
\leavevmode
\epsfysize=8.7cm
\epsfxsize=8.7cm
\epsffile{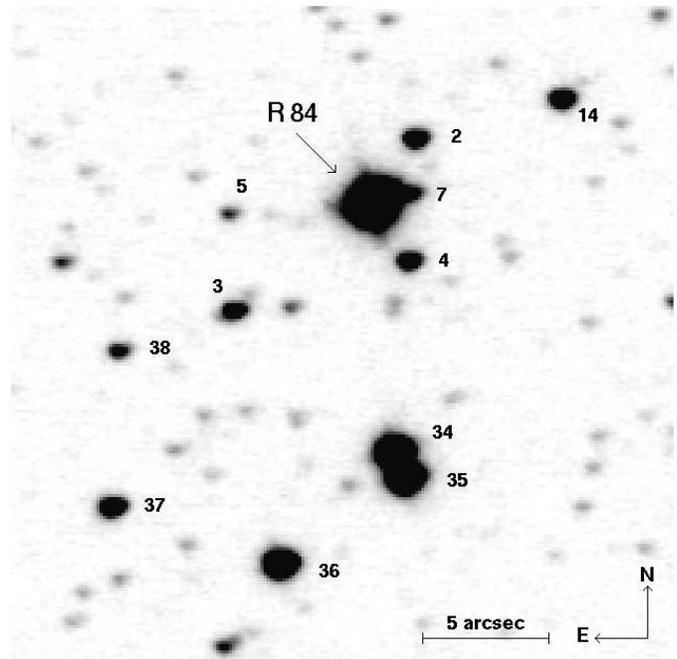}
\caption{An $R$ image of \R\, obtained using NTT+SUSI. Raw image with a 
   resolution of 0\frac.50\, ({\sc fwhm}). Field 
   $\sim$\,27\frac\,$\times$\,27\frac. Exposure time 1\,sec. 
North is at the 
   top and east to the left. Only the brighter components are labelled.}
\end{center}
\end{figure}

\parindent=1cm

Previously, NTT was  used on 1990 January 10, during the 
commissioning period of the telescope, when it was equipped with 
EFOSC2. The detector was a Tektronix
CCD (\#16) with 512$^{2}$ pixels of 27\,$\mu$m.   
The seeing varied between 0\frac.60 and 0\frac.75\, ({\sc fwhm}). 
Table 1 gives  more information on the images.

\begin{table*}
\caption[]{Journal of the imaging observations}
\begin{flushleft} 
\begin{tabular}{ c c c c c c c c }
\hline
Filter & ESO & \l    & $\Delta$\l & Expoure & Date & Pixel &
   Telescope + Instrument \\
    & (\#) & (\AA) & (\AA) &  (sec.) &      & (\arcsec) &            \\
\hline
$U$ & 631 & 3544.8 &  537.6 & 180.0 & 1990 Jan 10        & 0.146 &
   NTT\,+\,EFOSC2     \\
$B$ & 639 & 4340.8 & 1012.0 &   2.0 & 1991 Dec 26        & 0.128 &
   NTT\,+\,SUSI       \\ 
$B$ & 583 & 4399.8 &  941.9 &  15.0 & 1990 Jan 10        & 0.146 &
   NTT\,+\,EFOSC2     \\
$B$ & 445 & 4334.7 & 1017.0 &  30.0 & 1988 Aug 31        & 0.176 &
   2.2\,m\,+\,Adapter \\
$V$ & 641 & 5475.3 & 1130.2 &   1.0 & 1991 Dec 26        & 0.128 & 
   NTT\,+\,SUSI       \\
$V$ & 584 & 5476.3 & 1131.9 &  15.0 & 1990 Jan 10        & 0.146 &
   NTT\,+\,EFOSC2     \\
$V$ & 446 & 5446.5 & 1168.9 &  20.0 & 1988 Aug 31        & 0.176 &
   2.2\,m\,+\,Adapter \\
$R$ & 642 & 6437.2 & 1656.9 &   1.0 & 1991 Dec 26        & 0.128 &
   NTT\,+\,SUSI       \\
$R$ & 447 & 6483.8 & 1632.6 &  20.0 & 1988 Aug 31        & 0.176 & 
   2.2\,m\,+\,Adapter \\ 
$H$ &     &        &        & 12\,$\times$\,20.0 & 1995 Aug 20   
   & 0.050 & 3.6\,m\,+\,ADONIS  \\
$H$ &     &        &        & 100\,$\times$\,3.0 & 1995 Dec 31
   & 0.050 & 3.6\,m\,+\,ADONIS  \\
$K$ &     &        &        & 30\,$\times$\,20.0 & 1995 Aug 20  
   & 0.050 & 3.6\,m\,+\,ADONIS  \\
$K$ &     &        &        & 30\,$\times$\,20.0 & 1995 Dec 31 
    & 0.050 & 3.6\,m\,+\,ADONIS  \\
\hline
\vspace*{5 mm}
\end{tabular}
\end{flushleft}  
\end{table*}

\parindent=1cm

Additional observations were carried out on 1988 August 31 
at the ESO 2.2\,m telescope using the adapter for direct imaging. 
The detector was an RCA CCD chip (\#8) with 1024\,$\times$\,640 pixels 
of 15 $\mu$m size.  
The seeing conditions were poor, $\sim$\,1\frac.3\, ({\sc fwhm}). However,  
the comparison of these 2.2\,m observations with those obtained at 
the NTT telescope was very useful for checking the deconvolution 
code. 

\parindent=1cm

The data were all bias substracted and flat-fielded. Only the image 
in the $U$ band could not be flat-fielded, 
because of the too low S/N ratio of the flat-fields.  On the other hand, 
the SUSI CCD produced some non-Gaussian noise in the images but 
at a very low level, negligible at the S/N ratio of the stars 
studied here. 
Since \R\, is a very bright object, it was not always possible to avoid  
saturation, especially on our good seeing observations. These are 
the SUSI $R$ image which had 20 saturated pixels over \R, and 
the EFOSC2 $V$ image with 10 saturated pixels. We, therefore, used the 
observations of the other runs to  check the results.

\parindent=1cm

The  photometry of the  objects in the field  of  \R\, was carried
out with a new deconvolution algorithm  allowing not only to improve
the  spatial resolution of the  images, but also to obtain reliable
astrometric and photometric  measurements  of the stars. 
A full description of the method is given in Magain et al.\,(1997).
The principle of that method is to avoid deconvolving with the
total Point Spread Function (PSF), which would aim at obtaining
infinite resolution.  Rather, the new deconvolution allows to
obtain an image with a better (but not infinitely narrow) PSF,
basically chosen by the user.

\begin{table*}
\caption[]{Photometry of the core of LH\,39  on the basis 
of deconvolved images.\hspace*{.15cm}Star \#1 is \R} 
\begin{flushleft}
\begin{tabular}{c c c c c c c c c c c c}
\hline 
{\it Star} & $U$   &    $B$ & $B$ &  $B$   & $V$ & $V$ &  $V$ &  $R$ & 
   $R$  &  $x$ &   $y$  \\
\hline
 & EFOSC2 & EFOSC2 & 2.2\,m & SUSI & EFOSC2 &2.2\,m & SUSI & 2.2\,m & 
   SUSI & (\arcsec) & (\arcsec)\\
\hline		
1  & 11.33 & 12.26 & 12.26 & 12.26 & 12.10 & 12.10 & 12.10 & 11.58 & 
   11.58 &  0.00 &  0.00\\
2  & 15.29 & 16.20 & 16.09 & 16.00 & 16.33 & 16.24 & 16.04 & 16.21 & 
   15.99 & 1.75W & 2.67N\\  
3  & 15.29 & 16.13 &  $-$  & 16.06 & 16.27 &  $-$  & 16.24 &  $-$  & 
   16.28 &  5.60E &  4.36S\\
4  & 15.14 & 16.07 & 16.04 & 15.96 & 16.18 & 16.07 & 16.09 & 15.98 &
  16.07 & 1.49W &  2.33S\\
5  & 17.09 &  $-$  &  $-$  & 17.63 & 17.76 &  $-$  & 17.68 &  $-$  & 
   17.64 &  5.78E &  0.39S\\
6  &  $-$  &  $-$  &  $-$  &  $-$  &  $-$  &  $-$  &  $-$  &  $-$  & 
   18.61 & 0.93W &  3.98S\\
7  & 18.43 &  $-$  &  $-$  & 17.50 & 17.86 &  $-$  & 17.47 & 16.94 & 
   16.83 & 1.60W & 0.46N\\
8  &  $-$  &  $-$  &  $-$  &  $-$  &  $-$  &  $-$  &  $-$  &  $-$  & 
   18.81 & 5.60W &  2.16S\\
9  &  $-$  &  $-$  &  $-$  &  $-$  &  $-$  &  $-$  &  $-$  &  $-$  & 
   18.82 &  0.54E & 6.04N \\ 
10  &  $-$  &  $-$  &  $-$  & 18.50 &  $-$  &  $-$  & 18.45 &  $-$  & 
   18.41 & 0.61W & 7.55N \\
11  &  $-$  &  $-$  &  $-$  &  $-$  &  $-$  &  $-$  & 18.40 &  $-$  & 
   17.87 &  3.28E &  4.21S\\
13  &  $-$  &  $-$  &  $-$  &  $-$  &  $-$  &  $-$  &  $-$  &  $-$  &  
   $-$  & 3.16W &  3.16S\\
14  & 14.72 &  $-$  &  $-$  & 15.68 &  $-$  &  $-$  & 15.83 &  $-$  & 
   15.83 & 7.67W & 4.32N \\
15  &  $-$  &  $-$  &  $-$  &  $-$  &  $-$  &  $-$  &  $-$  &  $-$  &  
   $-$  & 2.27W & 1.50N \\
16  &  $-$  &  $-$  &  $-$  &  $-$  &  $-$  &  $-$  &  $-$  &  $-$  &  
   $-$  & 0.87W &  4.43S \\
19  &  $-$  &  $-$  &  $-$  &  $-$  &  $-$  &  $-$  &  $-$  &  $-$  &  
   $-$  &  4.73E & 2.58N \\
20  &  $-$  &  $-$  &  $-$  &  $-$  &  $-$  &  $-$  &  $-$  &  $-$  &  
   $-$  & 5.31W &  1.53S\\
21  & 16.64 &  $-$  &  $-$  & 16.20 &  $-$  &  $-$  & 16.72 &  $-$  & 
   16.66 & 0.59W & 0.87N \\
22  &  $-$  &  $-$  &  $-$  &  $-$  &  $-$  &  $-$  &  $-$  &  $-$  &  
   $-$  &  4.12E &  0.45S\\
23  &  $-$  &  $-$  &  $-$  &  $-$  &  $-$  &  $-$  &  $-$  &  $-$  &  
   $-$  & 8.79W &  4.84S \\
24  &  $-$  &  $-$  &  $-$  &  $-$  &  $-$  &  $-$  &  $-$  &  $-$  & 
   18.82 & 5.08W & 5.74N \\
25  &  $-$  &  $-$  &  $-$  &  $-$  &  $-$  &  $-$  &  $-$  &  $-$  & 
   18.93 &  1.46E & 7.33N \\
26  &  $-$  &  $-$  &  $-$  & 18.79 &  $-$  &  $-$  & 18.58 &  $-$  & 
   18.48 &  2.25E & 8.12N \\
27  &  $-$  &  $-$  &  $-$  &  $-$  &  $-$  &  $-$  &  $-$  &  $-$  & 
   18.96 &  4.96E &  3.68S\\
28  &  $-$  &  $-$  &  $-$  &  $-$  &  $-$  &  $-$  & 18.94 &  $-$  & 
   18.89 &  6.12E &  4.62S \\
29  &  $-$  &  $-$  &  $-$  &  $-$  &  $-$  &  $-$  &  $-$  &  $-$  &  
   $-$  & 8.57W & 5.02N \\
30  &  $-$  &  $-$  &  $-$  &  $-$  &  $-$  &  $-$  &  $-$  &  $-$  &  
   $-$  & 1.54W & 4.96N\\
31  &  $-$  &  $-$  &  $-$  &  $-$  &  $-$  &  $-$  &  $-$  &  $-$  &  
   $-$  & 3.28W & 4.36N \\
32  &  $-$  &  $-$  &  $-$  &  $-$  &  $-$  &  $-$  &  $-$  &  $-$  &  
   $-$  &  4.47E & 5.26N \\
33  &  $-$  &  $-$  &  $-$  &  $-$  &  $-$  &  $-$  &  $-$  &  $-$  &  
   $-$  &  2.81E &  0.73S \\
34  &  $-$  &  $-$  &  $-$  &  $-$  &  $-$  &  $-$  & 14.18 &  $-$  & 
   13.35 & 0.87W & 10.18S \\
35  &  $-$  &  $-$  &  $-$  &  $-$  &  $-$  &  $-$  & 13.45 &  $-$  & 
   13.38 & 1.32W & 11.21S \\
36  & 13.24 &  $-$  &  $-$  & 14.03 &  $-$  &  $-$  & 14.08 &  $-$  & 
   14.11 &  3.75E & 14.75S \\
37  & 17.95 &  $-$  &  $-$  & 17.15 &  $-$  &  $-$  & 16.58 &  $-$  & 
   15.92 & 10.56E & 12.43S \\ 
38  & 16.45 &  $-$  &  $-$  & 17.05 &  $-$  &  $-$  & 17.15 &  $-$  & 
   17.05 & 10.29E &  6.04S \\
\hline
\vspace*{5 mm}
\end{tabular}
\end{flushleft} 
\end{table*}

\begin{figure*}
%\picplace{8 cm}
%\begin{flushleft} 
\begin{center}
\leavevmode
%%\epsfysize=16.8cm
%%\epsfxsize=8.3cm
%%\epsffile{fig2hor.ps}
\epsfxsize=17.9cm
\epsffile{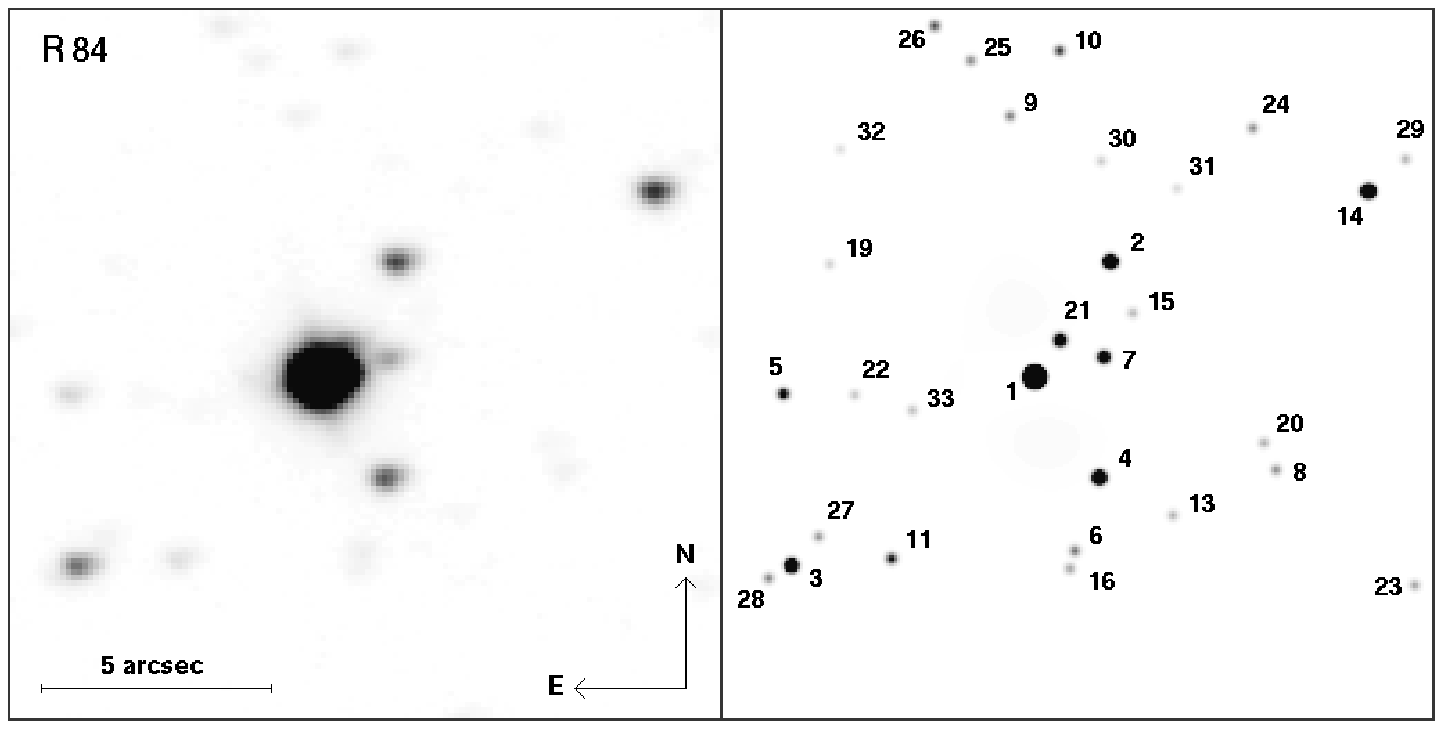}
\caption{A part of the  SUSI $R$ image focused on \R\, ({\it left}), and 
   the corresponding deconvolution result ({\it right}).     
   Star \#1 is the transition object \R. The pixel in the restored image is 
   two times smaller than in the  original one and the ``seeing'' is
   0\frac.19\,({\sc fwhm}). The faint diffuse  background in this image comes 
   from residual light due to diffraction   by   the  spikes of   the
   telescope. However, these residuals are negligible 
   compared with the
   intensity of the stars themselves (see the text). The intensity cuts of 
   this figure are chosen 
   to display the full dynamics of the image, even at low light levels.
   Field 16\frac.4\,$\times$\,16\frac.4. North is at the top and east 
   to the left.
 }
\end{center}
%\end{flushleft} 
\end{figure*}

\parindent=1cm

In the case of R84, the final PSF is chosen to be a Gaussian with
a FWHM of 3 pixels, the final pixel size being two times smaller than 
the original data pixels. 
The flux calibration was performed on the basis of the 
{\it UBVR} photometry carried out by Stahl et al (\cite{Sta84}) in 
August 1983 using a diaphragm of 15\frac\, in diameter. This is 
basically  the size of the field
we  use   for   the deconvolution. The integrated magnitudes  are therefore
re-distributed over all the components found. 

\parindent=1cm

Starting with a SUSI $R$ image of seeing 
0\frac.50 ({\sc fwhm)} presented in Fig. 1, we get 
a restored image of \R\, with a final resolution of 
0\frac.19 ({\sc fwhm}), which  is displayed in Fig. 2. 
We detect 31 components around \R\, over a 
$\sim$\,16\frac\,$\times$\,16\frac\, area. 
Owing to the high resolution of the images, for the first time 
we bring out stars \#2, \#4, and more especially \#21 and \#7 in the 
immediate vicinity of \R\, as well as the brighter components \#34 and \#35 
lying further away to the south. 
Among the stars for which we have color indices, 
there are three red stars, \R, \#7, and \#34. We will discuss 
about \R, and \#7 in Sect. 6. 
A prominent feature of \R\, is that  it turns up to be the 
reddest star of the field.  
The photometric and astrometric results are summarized in Table 2.
Note that the  magnitudes   of stars \# 34 to  \#38  were
obtained by aperture photometry.

Despite the  impression of  perfection  first  felt when looking at  the
deconvolved images,  one has to  remember  that it is   a model of  the
reality constructed  from imperfect data. 
If the PSF used  for the deconvolution is derived from stars as bright as
the object to deconvolve, Magain et al. (\cite{Mag97}) have shown that
the  photometry of the point sources is basically photon noise limited even
in the case of rather strong  blends (e.g. two stars as close  as one 
{\sc fwhm}). 
However,  in the  data  of  \R\, there are some additional error sources
in the astrometry and photometry: 
1) the PSF is constructed on stars at least five times fainter than \R;
2) \R\, itself is often saturated, sometimes heavily. 
Even  if for   most of  the  objects  in  the field  of \R\, the only
limitation  to the  photometric accuracy  is the  photon noise, the
effect of  an imperfect representation of  the   PSF  is not
negligible within a radius  of 1\frac\, of \R. 

\parindent=1cm

The PSF was constructed from 2 to  4 stars closer than 1\arcmin\, from \R,
in order to avoid any  possible PSF variation across the  field. In this
small area, no star as  bright as  \R\, is  available, especially in the red. 
In particular, the far wings of the PSF, as well as the diffraction 
spikes, are not modelled accurately enough for a perfect deconvolution 
of \R\, itself,
and this affects the photometry of the  closest neighbors, i.e.  stars
\#\,7  and \#\,21 (Fig. 2). 
Numerical simulations suggest that the uncertainty on the magnitudes
of star \#\,21 is of the order of 0.3 mag, while it amounts to 0.2
for star \#\,7, in  all the bands where we give a  magnitude for
these two objects. 
Note also that another consequence of the bad representation of the spikes
of the PSF is to produce a diffuse  background around  the bright
objects, especially \R. This halo (Fig.  2) is not real, but is 
neither an artefact due to the deconvolution algorithm. It is simply
due to  the
difference between the PSF used for the deconvolution and  \R\, itself. 
However, the relative  intensity between its highest values  and
the  faintest stars  is of the  order  of $10^{-2}$, negligible at the
precision we need for our purpose. 

\parindent=1cm

Anyhow, the PSF was accurate enough to allow 
the photometry of \R\, itself even from the frames where the 
star central pixels are 
saturated.  This was realized by giving an arbitrarily low weight
to the saturated pixels, so that the image of \R\, was modelled from the 
wings of its PSF.  Thanks to the good sampling of the original images,
this procedure gives an accurate estimate of the star's magnitude and
position.  This is confirmed by comparing the results with those obtained
from the unsaturated but  much lower  resolution  images taken with the 
ESO 2.2\,m telescope. Table 2 lists the magnitudes obtained for all 
the point sources with a S/N\,$>$\,10 in the   central pixel. 
The  typical error for a point  source with this S/N 
ratio is of the order of 0.1 magnitude.  

\subsection{Adaptive optics imaging and near IR photometry}

\R\, was observed in August and December 1995 with the 
ESO ADONIS adaptive optics system on the 3.6\,m telescope. Images were 
taken in the $H$ and $K$ bands with a pixel size of 0\frac.05.
For more details see Table 1. During the August run, four photometric
standards were also observed: HD\,115394, HD\,193901, HD\,207158
and HD\,218814, with the following exposure times~: 8\,$\times$\,15\,s,
20\,$\times$\,5\,s, 20\,$\times$\,5\,s and 4\,$\times$\,45 s in both 
$H$ and $K$. 
The reference star SAO\,249234 was also observed, for
later deconvolution, with an exposure time of 100\,$\times$\,3\,s
in both $H$ and $K$. 

\parindent=1cm

The images taken in August 1995 were affected by a strong noise
due to the poor quality of the detector during that run. 
Moreover, a very bad seeing (2 to 3\frac\,) throughout the night
was responsible for a very poor adaptive optics correction. 
For instance, the 
Strehl ratio varied between 0.008 and 0.1 in $K$ and the {\sc fwhm}
between 0\frac.6 and 1\frac.2. For these reasons, the observations
of August 1995 only showed the two brightest stars of the field 
(\R\, and \#11). 
These observations were nevertheless vital to perform a photometric 
calibration of the main star, thanks to the four photometric standards.
This was done using an aperture of diameter 5\frac. 
The transformation from the instrumental system to the  standard  
photometric system was carried out using the IRAF/NOAO PHOTCAL 
package. Note that the transformation was only possible
for the August data as no photometric
standard had been observed in December.

\parindent=1cm

The photometric calibration enabled us to calculate the magnitude of star \R: 
$H$\,=\,8.56\,$\pm$\,0.04,\,  $K$\,=\,8.13\,$\pm$\,0.03. 
These results agree very well with those of Stahl et al. (\cite{Sta84}). 
The errors  include an uncertainty due to variations
of the PSF with time estimated to be about 0.01
mag for our integration time (Esslinger \& Edmunds \cite{Ess}).
The results were also checked by performing the same
operation with some other aperture diameters.
Once we had the magnitude of \R, we could use it to
calibrate the images taken in December 1995. This was carried out
by measuring the flux of the star on these images with an aperture
of the same size as before.

\begin{table}
\label{tab:photom}
\caption{$H$ and $K$ photometry of the cluster} 
\begin{flushleft}
\begin{tabular}{c c c c }
\hline
{\it Star}  & $H$ & $K$ & {\it H\,--\,K} \\
\hline
1 &  8.56 &  8.13 & 0.43 \\
2 & 16.28 & 16.26 & 0.02 \\
4 & 16.05 & 16.08 & --0.03 \\
6 & 16.88 & 16.87 & 0.01 \\
7 & 14.38 & 14.16 & 0.22 \\ 
8 & 16.85 & 16.74 & 0.11 \\
11 & 15.64 & 15.50 & 0.14 \\
12 & 17.15 & 17.17 & --0.02 \\
15 & 16.78 & 16.59 & 0.19 \\
16 & 17.23 & 17.19 & 0.04 \\
17 & 16.98 & 17.00 & --0.02 \\
19 & 18.98 & 18.61 & 0.37 \\
22 & 17.11 & 16.74 & 0.37 \\ 
\hline
\end{tabular}
\end{flushleft}
\end{table}

\begin{figure}
%\picplace{8 cm}
\begin{center}
\leavevmode
\epsfysize=8.5cm
\epsfxsize=8.5cm
\epsffile{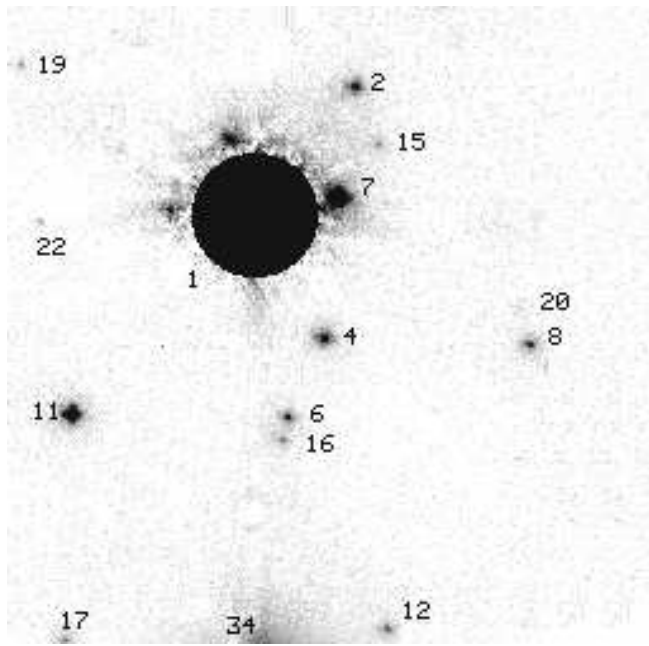}
\caption{An undeconvolved adaptive optics image of \R\, obtained using 
   ADONIS at 
   the ESO 3.6\,m telescope through an $H$ filter. Resolution 
   0\frac.12 ({\sc fwhm}) without deconvolution.  
   \R\, being comparably very bright, 
   we subtracted a properly placed and scaled reference star from  
   it and masked the residual
   in order to bring out the fainter stars. 
   Field 12\frac.8\,$\times$\,12\frac.8.
   North is at the top and east to the left.
 }
\end{center}
\end{figure}

\parindent=1cm

The images taken in December 1995 were very good 
(Strehl ratios 0.13 and 0.25 in $H$ and $K$, 
{\sc fwhm} resolutions 
0\frac.12 and 0\frac.15 respectively) and showed 17 stars in a 
field of 12\frac.8\,$\times$\,12\frac.8. 
To perform photometry we used both aperture photometry and 
PSF fitting in the IRAF/NOAO APPHOT and DAOPHOT
packages. We used an aperture of diameter 0\frac.5. 
This size was chosen to include
at least two dark rings of the diffraction-limited image, 
which limited the errors due to anisoplanatism to
less than 0.01 mag (Esslinger \& Edmunds \cite{Ess}).
PSF fitting, which is more sensitive to anisoplanatism, 
was only used to check the results of aperture photometry.  
Table 3 shows the results for each star, i.e. 
the magnitudes in $H$ and $K$ and the {\it H\,--\,K} colors. 

\parindent=1cm

The accuracies for stars other than \R\, were generally better than 
0.1 mag in both $H$ and $K$ bands. For star \#15, with $H$\,=\,16.74,  
it was about 0.3, and the worst was for star \#19 of $H$\,=\,18.94 
which amounted to 0.8 mag. These were 
mainly the errors given by the photometry packages.
For stars \#7 and \#15, which were in the halo of \#1,
the packages did not give accurate errors. We had to estimate them 
by changing the size of the aperture and checking the variations
in the fluxes. We found that the magnitudes of stars \#7 and \#15 
were respectively 
inaccurate within 0.2 and 0.3 mag in both bands. 
Stars \#12, \#17, and \#19 were  very close to the edge of the field
of view, especially in $K$. This forced us to take a smaller
aperture and also estimate the error ourselves. The magnitude of 
star \#34 was not measured, since only a part of its halo was visible.

\parindent=1cm 

In an attempt to deconvolve our images from the  run of December 1995,
we used the PSF calibration star.  Due  to  temporal
variation  of the  ADONIS PSF,  the  result  of  the  process is  very
disappointing.   Both simple algorithms  such  as  the Lucy-Richardson
method and our new algorithm leave  very significant residuals, due to
the fact that the PSF used for the deconvolution is not the actual PSF
of the image. As a result, we cannot detect faint objects very close
to   \R. For display purposes,  and  to enhance contrast  of the faint
objects, we subtracted the  PSF from \R. In  Fig.  3  we show the
result of the operation, where the strongest residuals are masked.

\subsection{Spectroscopy: CASPEC echelle and EMMI long slit}
 
\R\, was observed with the CASPEC spectrograph attached to the 3.6\,m
telescope on 1989 September 14. The 31.6 lines mm$^{-1}$ grating was used
with a 300 lines mm$^{-1}$ cross dispersion grating and an f/1.5 camera.
The detector was CCD\,\#8, a high resolution chip of type RCA SID\,006 EX with
1024\,$\times$\,640 pixels and a pixel size of 15\,$\mu$m. The central 
wavelength was \l4250\,\AA\, and the useful wavelength range \l\l3850 to 
4820\AA\, corresponding to orders 118 to 148 of the
Thorium-Argon calibration arc. The resulting {\sc fwhm} resolution as 
measured on the calibration lines is $\sim$\,0.2\,\AA. 
All the reductions were performed using the ECHELLE context of the 
MIDAS package. No flat-field correction was applied since the echelle 
orders in the flat-field frame appeared not aligned with the orders in 
the object frame. Orders 136 and 137 (\l\l4160 to 4190) are affected 
by a bad column of the detector and due to the lack of an appropriate 
flat-field correction, they could not be used. 
The individual orders were normalized by fitting fourth order polynomials 
and the accuracy of the normalization was checked by comparing 
overlapping regions of adjacent orders.
 
\parindent=1cm

Several moderate  resolution long-slit spectra
were taken of two stars (\#35 and \#36) 
in the direction of \R\, using NTT+EMMI with 
grating \#\,12 on 1993 September 22.
The CCD detector was 
Tektronix \#\,31 with 1024$^{2}$ pixels of size 24 $\mu$m.  The range was
\l\l3810--4740\,\AA\, and the dispersion  38\,\AA\,mm$^{-1}$, giving
{\sc fwhm} resolutions of $2.70\pm0.10$ pixels or $2.48\pm0.13$\,\AA\, for
a 1\frac.0 slit. 
Although the angular separation between stars 
\#35 and \#34 is only $\sim$\,1\frac, we expect no significant contamination 
of the spectrum of star \#35 by \#34 since the latter star is very faint 
in the blue. 

\begin{table*}
\caption[]{Lines measured in the spectrum of \R}
\begin{flushleft}
\begin{tabular}{ c c c c c c c c} \hline
\l & Ion & $v_{ab}$ & $v_{em}$ & Type & (EW)$_{ab}$ & (EW)$_{em}$ & $I_{em}$ \\
\hline
3888 & \hea + H8 & --133 & 232 & \p & 0.96 & --5.89 & 2.25\\
3934 & \ca & 246 & & A & & & \\
3965 & \hea & --65 & 200 & \p & & & \\
3968 & \ca & 248 & & A & & & \\
3970 & H$\epsilon$ & & 282 & E & & --1.79 & \\
3995 & \nb & & 238 & E & & --0.79 & \\
4026 & \hea & --38 & 290 & \p & 0.83 & --1.40 & 1.37\\
4089 & \sid & 61 & 238 & \p & & --0.64 & 1.28 \\
4097 & \nc & 16 & 236 & \p & & & \\
4101 & H$\delta$+\nc & & & E & & & 1.67 \\
4116 & \sid & & 213 & E & & --0.49 & 1.27 \\
4121 & \hea & & 277 & E & & --0.25 & 1.08 \\
4144 & \hea & & 267 & E & & --0.52 & 1.11 \\
4196 & \nc & & 207 & E & & & 1.06\\
4200 & \heb & (83) & & A & & & \\
4200 & \nc & & 259 & E & & & 1.06 \\
4340 & H$\gamma$\,+\,\nc & --125 & 260 & \p & 0.18 & --5.80 & \\
4340 & H$\gamma$\,(neb) & & 250 & E & & & \\
4379 & \nc & --6 & 223 & \p & 0.11 & --0.21 & 1.11 \\
4387 & \hea & --46 & 242 & \p & 0.06 & --0.94 & 1.24 \\  
4414 & [\fb] & & 230 & E & & & 1.05 \\
4416 & [\fb] & & 253 & E & & & 1.03 \\
4447 & \nb & & 226 & E & & --0.39 & 1.05\\
4471 & \hea & --66 & 293 & \p & 1.27 & --2.89 & 1.69 \\
4486$^*$ & ? & & & E & & --0.18 & 1.05 \\
4504$^*$ & ? & & & E & & --0.26 & 1.10 \\
4511 & \nc & 73 & 236 & \p & 0.12 & --0.14 & 1.08\\
4515 & \nc & (96) & 242 & \p & 0.08 & --0.18 & 1.09 \\
4518 & \nc & & 231 & E & & --0.08 & 1.02\\
4524 & \nc & & 227 & E & & & 1.01 \\
4534 & \nc & & 236 & E & & --0.15 & 1.07\\
4542 & \heb & 100 & & A & 0.22 & & \\
4552 & \sic & & 203 & E & & --0.32 & 1.08 \\
4568 & \sic & & 251 & \p & 0.33 & --0.38 & 1.08 \\
4634 & \nc & & 217 & E & & --1.13 & 1.46 \\
4640 & \nc & & 240 & E & & --1.64 & 1.58 \\
4647 & \cc & 71 & 221 & \p & 0.09 & & 1.11 \\
4651 & \cc & & 236 & E & & (--0.61) & 1.11 \\
4658 & [\fc] & & 253 & E & & --0.23 & 1.20 \\
4686 & \heb & & 222 & E & & --2.90 & 1.85 \\
4701 & [\fc] & & 259 & E & & --0.09 & 1.08 \\
4713 & \hea & --26 & 267 & \p & 0.58 & --0.84 & 1.20 \\
\end{tabular}
\end{flushleft}
\end{table*}

\begin{figure*}
%\picplace{8 cm}
\begin{center}
\leavevmode
\epsfysize=15.cm
\epsfxsize=15.6cm
\epsffile{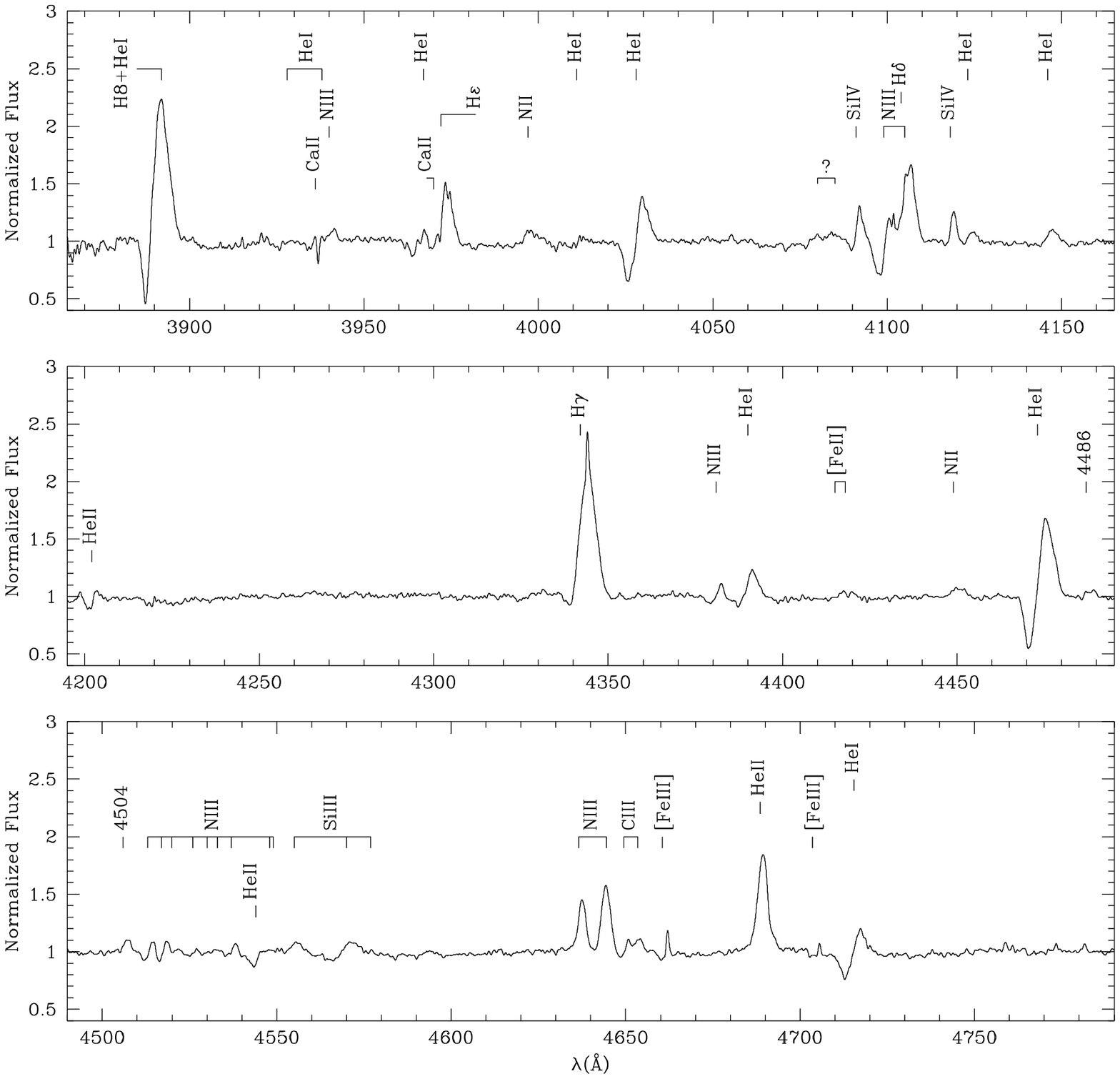}
\caption{Echelle spectra of \R\, obtained using CASPEC at the ESO 
   3.6\,m telescope in 1989. The normalized intensities of the strongest lines 
   are listed in Table 4.
 }
\end{center}
\end{figure*}

\begin{figure*}
%\picplace{8 cm}
\begin{center}
\leavevmode
\epsfysize=12.cm
\epsfxsize=15.cm
\epsffile{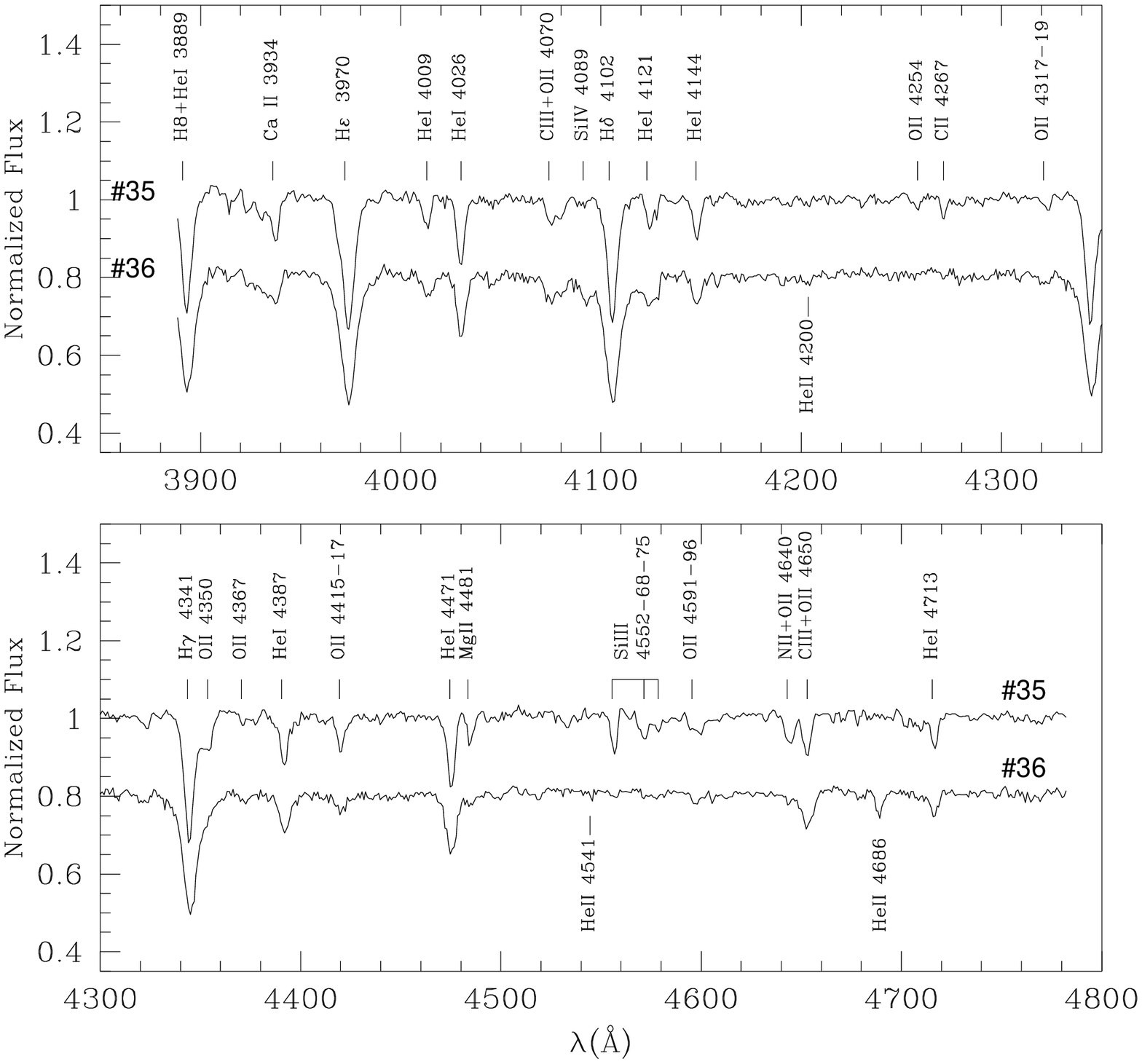}
\caption{Spectra of stars \#\,35 and \#\,36 obtained using the ESO 
   NTT+EMMI with grating \#12. 
 }
\end{center}
\end{figure*}

\section{Spectroscopy of \R} 

The fact that \O\, stars are transition 
objects linked to LBVs and so susceptible of variability, motivates 
monitoring of \R\, more especially as this star has shown photometric and 
spectroscopic variations. 
The CASPEC spectrum of \R\, obtained in 1989 September (JD 2447784.84) 
is displayed in Fig. 4. Besides the characteristic Of \heb\,\l4686 
and \nc\,\l\l4634-40 emission lines, the spectrum is dominated by strong 
emission lines of \ha, \hea, \nb, \nc, \sic\, and \sid, some of 
these lines displaying \p\, absorption components.

\parindent=1cm

Detailed information on our spectrum are listed in Table 4. Columns 3 
and 4 provide the heliocentric velocities of the absorption trough and 
emission peak respectively. The equivalent widths (EW) of the absorption 
components are listed in column 6, whereas columns 7 and 8 
respectively provide the EWs and normalized intensities of the emission 
components. The accuracies on the equivalent widths are $\sim 5\%$ for  
the stronger, non-blended lines and $\sim 10\%$ for weaker or blended 
features.

\parindent=1cm
We notice some differences in strength between the lines in our spectrum and 
the published data. Nota et al.\ (\cite{Not96}) report an EW of $-9.7$\AA\, 
for the H$\gamma$ + \nc\, emission line while we find a much lower value 
of $-5.8$\,\AA\, similar to the value of $-5.5$\AA\, measured by Crowther 
et al.\ (\cite{Cro95}) on the spectrum of Stahl et al.\ (\cite{Sta85}). 
The variability of this line was already noticed by Stahl et al.\ 
(\cite{Sta85}). Similar differences exist for the \hea\,\l\l4471 and 4713 
lines, that appear slightly stronger in Crowther et al.'s data and show EWs 
enhanced by nearly a factor of 2 in the Nota et al.'s spectra. 
In contrast, the 
absorption line \heb\,\l4542 has comparable intensities in the spectra 
obtained at different epochs. Although some differences in EW and intensity 
also exist for the Of-type emission lines, they are usually less severe than 
for the \ha\, and \hea\, lines. 
In general, the strongest variability is seen for those lines that 
are formed in the outer layers of the stellar wind. 
Note that IUE observations also indicate strong variations of the 
\nc\,\l1750 emission with EW between 1.5 and 5.3\,\AA\, 
(Hutchings \cite{Hut}). 

\parindent=1cm

Besides the well-known unidentified Of features at \l\l4486 and 4504, 
we find two unidentified weak and broad features at $\sim$\,\l\l4080 and 
4085\,\AA. The latter lines do not show up in previous observations reported 
by Stahl et al.\ (\cite{Sta85}), Wolf et al.\ (\cite{Wol87}), and  Smith et 
al.\ (\cite{Sm95}) and are out of the spectral range of Nota et al.\ 
(\cite{Not96}). These features are, however, seen on the spectra of Moffat 
(\cite{Mof89}) and could be marginally present in the spectra of Cowley \& 
Hutchings (\cite{Cow}) and Bohannan \& Walborn (\cite{Boh89}). Smith et 
al.\ (\cite{Sm95}) find a very broad emission feature in the region around 
the \heb\,\l4686, \nc\,\l\l4634-40 lines that they tentatively attribute 
to \nb. We find no such feature in our spectrum, but even if such a 
broad structure was present, it would cover the whole 
width of an order and would therefore be masked by the normalization 
procedure. As a consequence, depending on the adopted normalization, any 
measurement of the EWs of the \nc \l\l4634-40 and/or \heb \l4686 lines 
is more or less affected by this broad feature and any comparison of the 
strengths of these lines with data taken from the literature would be 
rather hazardeous. 

\parindent=1cm

Concerning the radial velocities of the \ha\, and \hea\, lines, 
the most important differences between 
our data and those of Nota et al.\ (\cite{Not96}) exist again for the 
\hea\,\l4471 (60\,km\,s$^{-1}$) and the \hea\,\l4713 (30\,km\,s$^{-1}$) 
lines. For most of the other lines, the differences are of the order of 
10--15\,km\,s$^{-1}$ which is about the estimated error of our measurements.

\parindent=1cm
 
Narrow nebular emission lines of [\fc] are detected at \l\l4658, 4701. 
A narrow emission component is also seen on top of the broad stellar 
H$\gamma$ emission component. These nebular features have widths of
 $50\pm2$\,\kms\, ({\sc fwhm}), in good agreement with the widths of 
the nebular H$\alpha$, H$\beta$ and [\nb] \l\l 6548,6583 lines (Nota 
et al.\ \cite{Not96}). The mean heliocentric radial velocity of the nebular 
lines in our spectrum is $254\pm4$\,\kms, in agreement with 
Cowley \& Hutchings's (\cite{Cow}) velocity of $256\pm1$\,\kms\, and 
Wolf et al.'s (\cite{Wol87}) value of 250 \kms\, both 
derived from the [\nb] lines.

\section{Stars \#\,35 and \#\,36}

The spectra of stars \#\,35 and \#\,36 obtained using long-slit 
spectroscopy with EMMI at the ESO NTT during excellent seeing 
conditions are presented in Fig. 5.   
The absence of \heb\, \l4686  in star \#35 indicates  a spectral 
type later than B0.5--B0.7  for this star of $V$\,=\,13.45 and 
{\it V\,--\,R}\,=\,0.07 (Walborn \& Fitzpatrick 
\cite{Wal90}). 
The line ratio \sic\,\l4552/\sid\,\l4089 being $>$\,1, implies a spectral 
type B1. The primary luminosity criterion at this type, i.e.  
\sic\,\l4552/\hea\,4387, which shows a smooth progression along the 
sequence (Walborn \& Fitzpatrick \cite{Wal90}), points to a supergiant 
class.  In fact, 
the spectral features of \#\,35 are reminiscent of those of  HD\,86606 
which is classified as B1\,Ib by Walborn \& Fitzpatrick (\cite{Wal90}). 
On the basis of  an interstellar reddening  
$A_{V}$\,=\,0.54 mag (Sect. 5) and a distance of 50 kpc (Westerlund 
\cite{Wes90}), an absolute visual magnitude $M_{V}$\,=\,--5.6 can be 
derived for 
star \#35, in good agreement with that expected for 
a B1\,Ib type (Lang \cite{Lan}). From measurement 
of 9 \hea\, and hydrogen lines we find a heliocentric radial velocity 
of 328\,$\pm$\,12 \kms\, for this star.

\parindent=1cm

Star \#\,36, being 0.63 mag weaker in $V$ than \#\,35, is  blue 
with colors {\it U\,--\,B}\,=\,--0.79, {\it B\,--\,V}\,=\,--0.05, 
and {\it V\,--\,R}\,=\,--0.03 mag. 
The presence of a  prominent \heb\,\l4686 line and very weak lines 
of \heb\,\l\l4200 and 4541 indicates  a very early-type B star. 
The ratio of \sic\,\l4552 to \sid\,\l4089 indicates a spectral type of B0.2. 
Walborn \& Fitzpatrick (\cite{Wal90}) noticed an extremely tight progression 
in the luminosity criteria at this spectral type. 
From the ratio of \sid\,\l4089 to \hea\,\l4121, which represents 
the main luminosity criterion, we deduce a main sequence class. 
Compared with  star \#\,35, we note significantly 
larger equivalent widths for \hea\, and  
broader \ha\, Balmer lines. These facts are also indicators of  
a main sequence luminosity class. 
However, assuming an intrinsic color of 
{\it (B\,--\,V)}$_{0}$\,=\,--0.30, we derive an absolute magnitude 
$M_{V}$\,=\,--5.1 which is too bright for a single 
main sequence early B type star. A medium resolution spectrum of star 
\#\,36 taken  with the Boller \& Chivens
spectrograph at the ESO 1.5\,m telescope in March 1997 
shows that most of the absorption
lines are double indicating that \#\,36 is a binary system with two stars of 
similar spectral types and luminosities. 

\parindent=1cm

Based on 10 \hea, \heb, and \ha\,  lines we derive a radial velocity of  
351\,$\pm$\,17 \kms\, for star \#36.

\section{Extinction towards the core of LH\,39}
  
The blue population of the color-magnitude diagram (Fig. 6) 
is centered on a color 
indice of {\it V\,--\,R}\,=\,0.05, {\it B\,--\,V}\,=\,--0.07 
(see also Table 2). 
Assuming an intrinsic color of 
{\it (B\,--\,V)}$_{0}$\,=\,--0.30 for O or early-type 
B stars  gives {\it E(B\,--\,V)}\,=\,0.23 mag. This corresponds to 
$A_{V}$\,=\,0.71 mag, 
or a reddening coefficient  of $c$(H$\beta$)\,=\,0.34, 
based on the interstellar extinction law derived by Savage \& 
Mathis (\cite{Sav79}) and $R$\,=\,3.1. This result is in good 
agreement with Lucke's (\cite{Luc}) photometry   
who finds a mean {\it E(B\,--\,V)}\,=\,0.18\,$\pm$\,0.09 (s.d.)
for a sample of 14 stars belonging to the OB association LH\,39 
(Schild \cite{Sch}).  

\parindent=1cm

The extinctions that we derive for stars \#35 and \#36 also fully agree
with this mean value. As to \#35, 
using an intrinsic color of {\it V\,--\,R}\,=\,--0.07 for a  
type B1\,Ib (Johnson, 
\cite{Joh68}), gives {\it E(V\,--\,R)}\,=\,0.14 mag, which corresponds  
to an interstellar reddening of {\it E(B\,--\,V)}\,=\,0.18 or 
$A_{V}$\,=\,0.54. For star \#36 
we find {\it E(B\,--\,V)}\,=\,0.24, or $A_{V}$\,=\,0.74 mag, 
using an intrinsic color indice $(B - V)_{0} = -0.29$.

\parindent=1cm

Regarding \R, its {\it B\,--\,V} color varied between 0.11 and 0.16 mag 
over a lapse of 15 nights from December 28, 1981 to January 11, 1982 (Stahl et 
al. \cite{Sta84}). Different values ranging from 0.04 to 0.15 were 
suggested for the {\it B\,--\,V} color excess of \R\, according to 
observations conducted at different periods between 1960 and 1990 
(Feast et al.\ \cite{Fea}, Stahl et al. \cite{Sta84}, 
Vacca \& Torres-Dodgen \cite{Vac90}, Crowther et al. \cite{Cro95}).   
Using Stahl et al.'s (\cite{Sta84}) Aug.\ 1983 observations, we derive 
{\it B\,--\,V}\,=\,0.16 for \R. 
Comparing UV, visible, 
and near IR fluxes, Stahl et al. (\cite{Sta84}) concluded that the 
late-type star contributes significantly to the flux in the $V$ band. 
Therefore, one cannot directly compare the observed colors 
with the intrinsic {\it (B\,--\,V)}$_{0}$ of --0.30 derived by Crowther 
et al.\ (\cite{Cro95}) for the W-R star in \R. Instead, we use 
the measured {\it U\,--\,B}\,=\,--0.93 (Table 2) to derive the 
reddening of \R. 
We estimated the intrinsic color {\it (U\,--\,B)}$_{0}$ from the 
photometric data for a sample of  7 LMC Ofpe/WN9 stars 
(Nota et al. \cite{Not96}). 
Using Savage \& Mathis's (\cite{Sav79}) interstellar extinction curve, 
the measured {\it B\,--\,V} colors, 
and {\it (B\,--\,V)}$_{0}$\,=\,--0.30 for these stars, we work out the 
color excess {\it E(U\,--\,B)}.  Then from the listed {\it U\,--\,B} 
data we get a mean intrinsic color 
{\it (U\,--\,B)}$_{0}$\,=\,--1.09\,$\pm$\,0.07 for the sample.   
This parameter is then used  to derive   
{\it E(B\,--\,V)}\,=\,0.24 for \R, which corresponds to 
$A_{V}$\,=\,0.75 mag,  in good agreement  with the extinction derived 
above for the members of LH\,39.

\section{Discussions}

\subsection{A word of caution about the colors}

The powerful deconvolution method decribed in Sect. 2.1 has allowed us 
to resolve the stars populating the core of the LMC OB association LH\,39 
in more than 30 components. More especially, it has shown the presence of two 
previously 
unknown stars, \#7 and \#21, which are the closest to \R\, at a resolution 
of 0\frac.19 ({\sc fwhm}) in the  optical domain. Furthermore, this 
deconvolution code has enabled us for the first time to perform the  
photometry of the components.  However, 
we should underline that this photometry is relative for 
a number of reasons which have nothing to do with the limitations of 
the code. We have used Stahl et al.'s (\cite{Sta84})  observations  
of \R\, to calibrate our {\it UBVR}  
observations carried out at epochs different from theirs. 
Moreover, since the calibration is based on only one ``standard star'', 
color corrections have not been possible. This shortcoming particularly 
affects the colors, and we have therefore  taken care not to 
over-interpret them. Another point, as discussed in Sect. 5, is the 
slight color variability of \R.  
However, this variability, of the order of 0.05 mag, is smaller than  
our inaccuracies. We envisage high resolution photometric observations 
including standard stars to 
improve the colors and use them for further study of \R.

\subsection{The late-type component}

\R\, is unique since it shows the features of both a transition \O\, type 
star and a late supergiant. No other star of this family is 
known to be associated with an evolved late-type star. 
The very large IR excess observed towards \R\, was attributed by 
Allen \& Glass (\cite{All76}) to  
the presence of a late-type supergiant component that provides 
the near-IR flux.  This component was 
spectroscopically detected by Cowley \& Hutchings (\cite{Cow}) who 
classified it as M2 on the basis of the relative strengths of 
TiO bands at \l\l5167, 5448, and 6159. Later, Wolf et al. (\cite{Wol87}) 
published a CASPEC spectrum containing several absorption lines mainly 
of neutral elements and TiO bands and confirmed the spectral 
type M2\,Ia. Also, McGregor et al. 
(\cite{McG88}) interpreted the CO absorption bands in the 2.0--2.4\,$\mu$m 
region of the spectrum of \R\, to be due to a cool supergiant 
companion star.

\parindent=1cm

Cowley \& Hutchings (\cite{Cow}), using a $K$ magnitude of 
8.49 (Allen \& Glass \cite{All76}) and 
Johnson's  calibration for M0 stars, estimated 
an absolute magnitude of 
$M_{V}$\,=\,--6.7 for the supergiant component. However,   
this is certainly an overestimate, since it assumes that the W-R  
star has no important contribution to the $K$ flux. 
Recent works on Ofpe/WN9 stars, LBVs, and related objects have shown 
the presence of extended envelopes around these stars that produce a 
strong IR excess in their spectrum.
Moreover, in the particular case of \R, Stahl et al. (\cite {Sta84}) 
provide evidence for the existence of a circumstellar dust shell 
especially on the basis of a large {\it K\,--\,L} color that cannot be 
explained by a late-type star alone.

From an absolute magnitude of $M_{V}$\,=\,--5.6, 
expected for an M2\,Iab star (Lang \cite{Lan}) and assuming that 
the putative late-type supergiant is a foreground star, 
we derive a $V$ magnitude of 13.2. This would be the brightest 
star in the field of view after \R. However, no star as bright is 
disclosed by our photometry using  resolutions of 0\frac.12 
(in $H$ band) and 0\frac.19  (in $V$) in a field of 
$\sim$\,15\frac\, centered on \R. 
The second brightest star of Table 2 with red colors, i.e. \#34 with 
$V$\,=\,14.18, is too far apart to fall into the 
large apertures  (diameters of order 15\frac\,) used in the classical 
photometry or to contribute to the spectrum of \R. 
The best candidate for the late-type companion seems 
therefore to be 
star \#7, lying at 0\frac.46 N, 1\frac.60 W of \R. It is the brightest and   
the  reddest star of the near-IR sample  
 and also the closest to \R\, (Table 3, Figs. 2, 3, 6).   
Moreover, its {\it H\,--\,K}\,=\,0.22 mag, although suffering from 
a rather large uncertainty, is compatible with that 
expected for an M2 type (Koornneef \cite{Koo}). 
However, with its 
$V$\,=\,17.47\,$\pm$\,0.2\, or $K$\,=\,14.10\,$\pm$\,0.22, it is extremely 
weak, unless it varies strongly. 
We will see below 
(Sect. 6.3) that  star \#7 is probably a line-of-sight object 
rather than a genuine association member. 

\parindent=1cm

\R\, is known to 
be variable. Its $V$ brightness has been reported to change by 0.2 mag 
between 1972 and 
1984 (see Stahl et al. {\cite{Sta84} and references therein). However, 
we do not know whether there is an offset in the zero points 
used by various observers.  Anyhow, Stahl et al. ({\cite{Sta84}), 
using a homogeneous set  of observations, found a magnitude variation 
of $V$\,=\,0.08 between 1983 and 1984. 
The variations in the near-IR magnitudes are also comparable to those in 
the optical domain (Allen \& Glass \cite{All76}, Stahl et al. 
\cite{Sta84}). 
From the similarity of the color variations of \R\, with those of 
R\,85 and R\,99 Stahl et al. (\cite {Sta84}) suggest that  they might 
be due to the blue star. However, since in the reported classical 
photometries rather large apertures are used (15\frac\,) we cannot 
{\it a priori} exclude 
the possibility that 
this similarity may be a coincidence and that part of the observed 
variations may be due to the other cluster stars falling in the 
aperture. However, in order to get a variation of 0.08 mag in 
the integrated $V$ magnitude, star \#7 or another star of the 
field should undergo  rather 
unrealistic variations.

\parindent=1cm

Although we cannot preclude the presence of a line of sight 
companion lying closer than 0\frac.12 to \R, the possibility of a 
binary system as suggested by McGregor et al.\ (\cite{McG88}) 
is very appealing. In order to investigate the 
possibility of a binary system, one has to 
measure the actual radial velocity of \R. However, this is not an easy task, 
since Crowther et al.\ (\cite{Cro95}) have shown that even the  
\heb\,\l4542 absorption is not a pure photospheric line, but is already 
affected by the stellar wind. Therefore, one has to rely on the radial 
velocities of pure emission lines. 
Cowley \& Hutchings (\cite{Cow}) 
measured a  radial velocity of $212\pm17$\,km\,s$^{-1}$ for the \nc, \sid, 
\heb\, and \cc\, emission lines on their spectrum obtained in Nov.\ 1977.
Moffat (\cite{Mof89}) found no indication 
of variations on time scales from a day to a year in his \heb\,\l4686 data 
obtained in 1978 and 1980. He derived a mean radial velocity of 
208 \,km\,s$^{-1}$ with a standard deviation of 5\,km\,s$^{-1}$. 
Nota et al.\ (\cite{Not96}) measured a radial 
velocity of 235\,km\,s$^{-1}$ in 1991 September, whereas our spectrum 
obtained in 1989 September yields a value of 222 km\,s$^{-1}$. Considering 
the mean radial velocity of the Of emission lines \heb\,\l4686 and 
\nc\,\l\l4634-40, we 
derive a mean velocity of 205\,$\pm 7$\,\kms\, from 
Moffat's (\cite{Mof89}) data set, whereas our spectrum yields a value of 
226\,$\pm$\,10 km\,s$^{-1}$ and Nota et al.\ (\cite{Not96}) 
find 246\,$\pm$\,13 km\,s$^{-1}$ 
including also the H$\alpha$ and \hea\,\l6678 lines in their mean. 
We have also measured the radial velocities on the AAT spectrum of 
Smith et al.\ (\cite{Sm95}) kindly provided by Dr.\ P.\ Crowther. 
We obtain a velocity of 269\,\kms\, for the \heb\,\l4686 line and 
256\,\kms\, for  the mean of the 
Of emission lines. These data are too scarce to draw 
any firm conclusion on the binary status 
of \R, but they clearly indicate the presence of radial velocity 
variations and are 
not incompatible with the possibility of a long-period low-inclination 
orbit.

\begin{figure}
%\picplace{8 cm}
\begin{center}
\leavevmode
\epsfysize=8.7cm
\epsfxsize=8.7cm
\epsffile{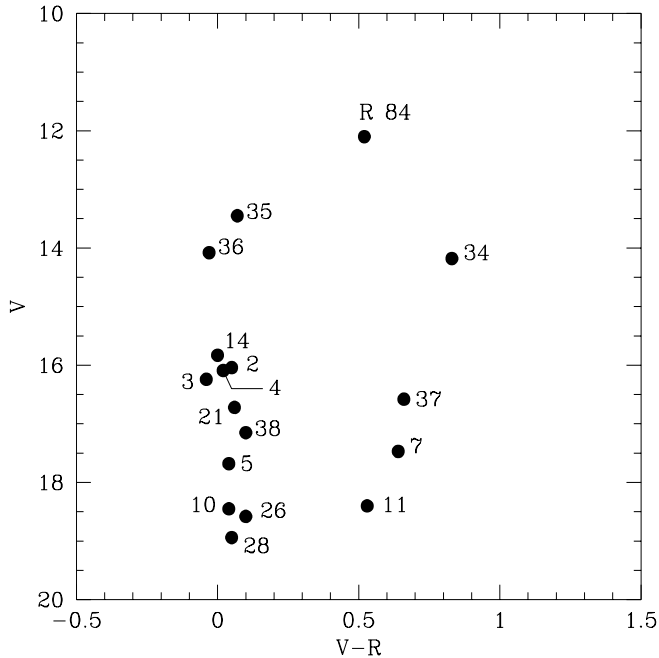}
\caption{Color-magnitude diagram of the components towards the core 
   of the OB association LH\,39 resolved by  deconvolution.  }
\end{center}
\end{figure}

\subsection{Cluster membership}

The color-magnitude diagram in the {\it V\,--\,R}, $V$ plane for 
stars with the deconvolution photometry is shown in Fig. 6. Two distinct 
groupings appear, a vertical, blue one lying at 
$<${\it V\,--\,R}$>$\,$\sim$\,0.1 and a smaller, red population 
with {\it V\,--\,R}\,$\geq$\,0.5 mag. 
However, the apparent ``main sequence'' may be contaminated by evolved, 
foreground stars less  affected by reddening.  
We must therefore check the effects of reddening upon the abscissa. 
A two-color {\it U\,--\,B}, {\it B\,--\,V}  diagram turns out to be a useful 
complement to separate the foreground stars from the association members. 
Unfortunately, in 
the field of \R\,  we have  only 11 stars with measured 
$U$, $B$, and $V$ photometry. Most of them appear to have a reddening-free 
index $Q = (U-B)-0.72(B-V)$ (Massey et al.\ \cite{Mass}) between $\sim 
-0.45$ and $\sim -0.85$ and lie on reddening lines between spectral types 
B5 and B0. These stars (\#2, 3, 4, 5, 14, 36, 38) are therefore very likely 
members of the OB association. However, photometry alone is not sufficient 
and as long as spectral types are not derived from spectroscopy we should 
be very cautious about these results. Three stars, \#7, \#21, and \#37, 
stand out of the main group. These  stars are subject to large
uncertainties (Sect. 2.1.3). Star \#37 is particularly  faint
in  $U$ and the fact that the $U$ image  could not  be flat-fielded 
introduced a large error in this band.   
The $3\sigma$ error bars on the 
colors of these stars are therefore 0.6, 0.4, 
and 0.3 mag respectively. 
Whereas star \#37 is most likely a red foreground star, 
the situation is less clear for star \#21 for which more accurate photometry 
is needed. The same holds true for star \#7, although this star is more 
likely a  line-of-sight component.

\begin{figure}
%\picplace{8 cm}
\begin{center}
\leavevmode
\epsfysize=8.7cm
\epsfxsize=8.7cm
\epsffile{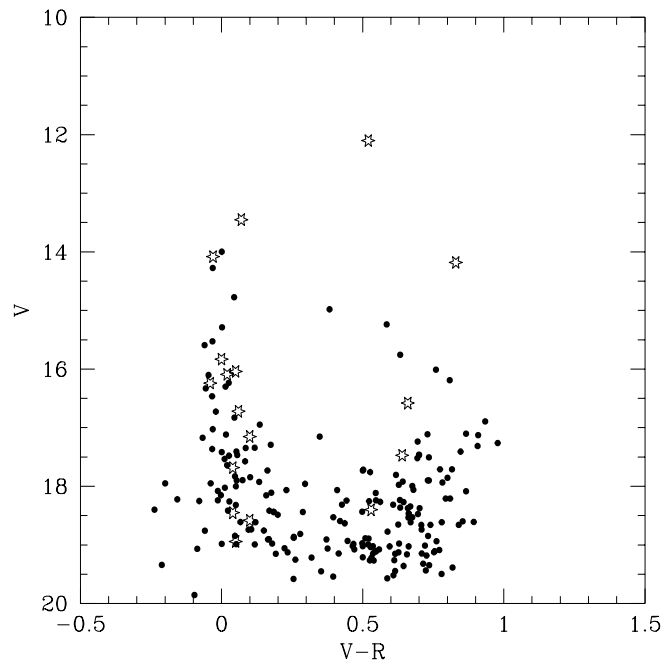}
\caption{Color-magnitude diagram of a total of 212 stars in the 
full 1\min.5\,$\times$\,1\min.8 SUSI field. The asterisks 
represent the deconvolution magnitudes 
(16\frac.4\,$\times$\,16\frac.4 crowded field  around \R)
and the black dots are the rather isolated stars for which we get 
aperture photometry.}
\end{center}
\end{figure}

Fig. 6 represents a small group of stars concentrated in the core 
of the association, whereas the color-magnitude diagram of the full 
SUSI field is displayed in Fig. 7. The magnitudes were obtained using 
S-Extractor, an efficient package 
for aperture photometry of large numbers of sources in large fields 
(Bertin \& Arnouts \cite{Ber}). The  code has allowed us to derive the 
photometry 
of the brightest, non-blended stars for comparison  with the results 
obtained using deconvolution in the smaller field. 
Since  \R\, is  saturated in  the $R$ filter, the flux
calibration in this band was done using the magnitude of the 
relatively bright star \#14, given by deconvolution (Table 2), 
whereas the $V$ band results are 
based on the magnitude of \R. 
A remarkable feature of the color-magnitude diagram is the good 
agreement between the characteristics of the different groupings obtained by 
the deconvolution code applied to the crowded central part of the field on 
the one side and the aperture method applied to isolated stars in the whole 
SUSI field on the other side. Whereas the sequence with $V-R\sim 0.05$ 
essentially traces the OB association, the redder group  with $V-R>0.5$ 
is due to LMC field stars. This field star grouping may be a mix of several 
populations, as Walker (\cite{Walk}) has shown in the case of the LMC 
cluster NGC\,1866.

\subsection{Radial velocities}

We have measured the radial velocities of 15 known members of the 
LH\,39 association (Schild \cite{Sch}) on the spectra kindly provided 
by Dr.\ H.\ Schild. Since these spectra have a low {\sc fwhm} resolution of 
6.8\,\AA, they can only provide a rough estimate of the radial velocities. 
Our measurements span a rather wide range between 215 and 440\,km\,s$^{-1}$. 
The above derived radial velocities of stars \#35 and \#36 are therefore 
fully consistent with the mean value of the LH\,39 association, whereas the 
velocity of \R\, lies towards the lower end of our range.

\section{Concluding remarks}

The magnitudes of \R\, reported in the literature are based on 
classical photometry using large apertures of some 15\frac\, in diameter.  
However, this work has shown that about 30 stars populate these 
apertures and the derived magnitudes are overestimates although 
\R\, dominates the cluster. In deriving the magnitudes of \R\, we have 
surpassed this crowding problem.  On the other hand, 
if the late-type star is confirmed to be a  line-of-sight companion,  
this means that  
the visual magnitudes so far given for \R, including our relatively 
lower values (Table 2), are all upper limits. Assuming an apparent 
magnitude 
of $V$\,=\,13.2 for the M2 supergiant (see Sect. 6.2) leads to a $V$\,=\,12.6 
for \R, based on  a global magnitude of 12.10 (Table 2) for both stars.  
This corresponds to an absolute magnitude of $M_{V}$\,=\,--6.6, 
which is fainter than the recently derived value of --7.0 
(Crowther et al. \cite{Cro95}).

A remarkable point is the apparent absence of O type stars in the 
OB association LH\,39. Schild (\cite{Sch}) derived the spectral types of 
the 16 brightest blue stars of the association. In this sample 13 stars 
are B types (only 3 of them main sequence), there are 2 A supergiant  
stars, and \R. Stars \#35 and \#36 that we have observed towards the 
core of the association are also B types (Sect. 4). One of them may 
be a double system with spectroscopically identical components. 
Several other stars in the association, for which no spectroscopy is 
available, have a reddening-free $Q$ index consistent with a B type 
classification (Sect. 6.3). For instance, stars \#3 and \#14 
have $V$\,=\,16.24 and 15.83 (Table 2) corresponding 
to absolute visual magnitudes of about --3.0 and --3.5 respectively. 
These are too faint for O types and would rather indicate 
main sequence B stars. 

\parindent=1cm

If there were any undetected O stars in this association, they should be 
fainter than the main sequence B types, and this is difficult to admit 
since the reddening  is pretty uniform over the association and we do not 
expect important local extinction in this rather evolved association. 
Therefore, we conclude that the turn-off of this association lies around B0. 
This entails a rather provocative question as to the progenitor of \R. 
Could \R\, come from a B type star? Of course one is inclined to 
admit that the progenitor of \R\, was the most massive star of the cluster, 
a supposedly O type. Generally speaking, the formation of O type stars is a 
collective process, as indicated  by recent observational and theoretical 
works (see Heydari-Malayeri \cite{Hey} and references therein). But there is 
no observational evidence supporting the existence of O 
stars in LH\,39, 
as is the case in other LMC OB associations (e.g. LH\,117 and LH\,118, 
Massey et al. \cite{Mas}).

According to the Z\,=\,0.008 evolutionary tracks of Schaerer et al.\ 
(\cite{Schaerer}), the mass limit capable of  forming a W-R star is slightly 
above 40\,\sm\, and \R\, lies very close to the 40 \sm\, track. 
Although, very massive Galactic late-type WN stars are 
probably still in the H-burning stage (Rauw et al.\ \cite{Rauw}), it seems 
unlikely that less massive stars in the LMC will reach the W-R stage prior 
to the He-burning phase. Therefore, the total H\,+\,He-burning lifetimes 
for a 40\,\sm\,  
star will set an upper limit of 5.3 Myr to the age of \R, whereas most of 
the association members are consistent with an age of about 12 Myr. 
Unless the mass limit to form a W-R star is considerably lower than 40\,\sm, 
we are forced to admit that the stars in LH\,39 are not coeval. 

\parindent=1cm

The presence of a low mass M supergiant in this OB association is another 
challenging problem since Humphreys et al.\ (\cite{HNM}) showed that red 
supergiants and W-R stars were anticorrelated in the OB associations of 
M\,33 (metallicity similar to the LMC). Its formation might be more 
easily understandable if it belongs to a binary system where the physical 
effects of mass exchange would considerably alter the evolutionary paths. 
If the partner is effectively the Ofpe/WN9 star, such a scenario could also 
provide an explanation for the present status of \R.

\acknowledgements{We would like to thank 
Dr. Hans Schild for providing his spectra of 
16 members of the LH\,39 association.       
We are grateful to Dr. Paul Crowther for supplying the AAT spectrum 
of \R. We are also indebted to   
Dr. Jean-Luc Beuzit for his 
assistance during the adaptive optics observations and Dr. Laurent Jorda for 
a preliminary reduction of the NTT+EFOSC2 data.   
O. Esslinger is supported by a PPARC PhD grant through the United
Kingdom Adaptive Optics Program. 
G. Rauw, P. Magain, and F. Courbin acknowledge financial support from 
contract ARC\,94/99-178 ``Action de recherche concert\'ee
de la Communaut\'e Fran\c{c}aise'' (Belgium). G. Rauw also 
acknowledges support from the Fonds National de la Recherche 
Scientifique (Belgium) and partial support through 
the PRODEX XMM-OM Project. 
M. Heydari-Malayeri recognizes financial help from the 
Institut National des Sciences de l'Univers (INSU) to participate in the 
workshop ``Luminous Blue Variables: Massive Stars in Transition'',  
Kona, October 6--11, 1996. 
The SIMBAD database was  
consulted for the bibliography.
}


\begin{thebibliography}{}


\bibitem[1976]{All76}
Allen, D.A., Glass, I.S., 1976, ApJ 210, 666 
\bibitem[1996]{Ber}
Bertin, E., Arnouts, S., 1996, A\&AS 117, 393
\bibitem[1989]{Boh89}
Bohannan B., Walborn N.R., 1989, PASP 101, 520 
\bibitem[1981]{Br81}
Breysacher, J., 1981, A\&ASS 43, 203 
\bibitem[1978]{Cow}
Cowley, A.P., Hutchings, J.B., 1978, PASP 90, 636
\bibitem[1995]{Cro95}
Crowther, P.A., D.J. Hillier, Smith, L.J., 1995, A\&A 293, 172 
\bibitem[1976]{Dav}
Davies, R.D., Elliott, K.H., Meaburn, J., 1976, MNRAS 81, 89 
\bibitem[1997]{Ess}
Esslinger, O., Edmunds, M.G., 1997, A\&A, in preparation
\bibitem[1960]{Fea}
Feast, M.W., Thackeray, A.D., Wesselink, A.J., 1960, MNRAS 121, 337
\bibitem[1956]{Hen56}
Henize, K.G., 1956, ApJS 2, 315
\bibitem[1996]{Hey}
Heydari-Malayeri, M., 1996, Interplay between Massive Star Formation , 
   the ISM, and the Galaxy Evolution, D. Kunth et al. (eds), Editions 
   Fronti\`eres, p. 51
\bibitem[1984]{HNM}
Humphreys, R.M., Nichols, M., Massey, P., 1984, AJ, 90, 101 
\bibitem[1980]{Hut}
Hutchings, J.B., 1980, ApJ 237, 285  
\bibitem[1968]{Joh68}
Johnson, H.L., 1968, in {\it Nebulae and Interstellar Matter}, 
   B.M. Middlehurst \& L.H. Aller (eds.), Chicago Press, p. 167  
\bibitem[1983]{Koo}
Koornneef, J., 1983, A\&A 128,84
\bibitem[1992]{Lan}
Lang, K.R., 1992, {\it Astrophysical Data: Planets and Stars}, Springer-Verlag
\bibitem[1994]{Langer}
Langer, N., Hamann, W.-R., Lennon, M., Najarro, F., Pauldrach, A.W.A., 
Puls, J., 1994, A\&A, 290, 819
\bibitem[1972]{Luc}
Lucke, P.B., 1972, Thesis, University of Washington
\bibitem[1970]{LH}
Lucke, P.B., Hodge, P.W., 1970, AJ 75, 171
\bibitem[1989]{Maeder}
Maeder, A., 1989 in Proc.\ IAU Coll.\ 113, Physics of Luminous Blue 
   Variables, K.\ Davidson, A.F.J.\ Moffat \& H.J.G.L.M.\ Lamers (eds.), 
   Kluwer, p.\,15
\bibitem[1997]{Mag97}
Magain, P., Courbin, F., Sohy, S., 1997, submitted to ApJ 
   (SISSA preprint astro-ph/9704059)
\bibitem[1989]{Mas}
Massey, P., Garmany, C.D., Silkey, M., Degioia-Eastwood, K., 1989, 
   AJ 97, 107
\bibitem[1995]{Mass}
Massey, P., Lang, C.C., DeGioia-Eastwood, K., Garmany, C.D, 1995, ApJ 438, 188
\bibitem[1988]{McG88}
McGregor, P.J., Hillier, D.J., Hyland, A.R., 1988, ApJ 334, 639
\bibitem[1989]{Mof89}
Moffat, A.F.J., 1989, ApJ 347, 373 
\bibitem[1996]{Not96}
Nota A., Pasquali A., Drissen L., Leitherer C., Robert C., 
   Moffat A.F.J., Schmutz W., 1996, ApJS 102, 383 
\bibitem[1996]{Rauw}
Rauw, G., Vreux, J.-M., Gosset, E., Hutsem\'ekers, D., Magain, P., Rochowicz, 
   K., 1996, A\&A 306, 771  
\bibitem[1970]{San70}
Sanduleak, N., 1970, in {\it Cerro Tololo Inter-American Obs. Cont. 89} 
\bibitem[1979]{Sav79}
Savage, B.D., Mathis, J.S., 1979, ARA\&A 17, 73
\bibitem[1993]{Schaerer}
Schaerer, D., Meynet, G., Maeder, A., Schaller, G., 1993, A\&ASS 98, 523 
\bibitem[1987]{Sch}
Schild, H., 1987, A\&A 173, 405 
\bibitem[1991]{Sch91}
Schmutz, W., Leitherer, C., Hubeny, I., Vogel, M.,  
   Hamann, W.-R., Wessolowski, U., 1991, ApJ 372, 664 
\bibitem[1968]{Smi68}
Smith, L.F., 1968, MNRAS 140, 409 
\bibitem[1994]{Sm94}
Smith, L.J., Crowther, P.A., Prinja, R.K., 1994, A\&A 281,  833 
 \bibitem[1995]{Sm95}
Smith, L.J., Crowther, P.A., Willis, A.J., 1995, A\&A 302, 830
\bibitem[1983]{Sta83}
Stahl, O., Wolf, B., Klare, G., Cassatella, A., Krautter, J., 
   Persi, P., Ferrari-Toniolo, M., 1983, A\&A 127, 49 
\bibitem[1984]{Sta84}
Stahl, O., Wolf, B., Leitherer, C., Zickgraf, F.-J., Krautter, J., 
   de Groot, M., 1984, A\&A 140, 459 
\bibitem[1985]{Sta85}
Stahl, O., Wolf, B., de Groot, M., Leitherer, C., 1985, A\&ASS 61, 237  
\bibitem[1990]{Vac90}
Vacca, W.D., Torres-Dodgen, A.V., 1990, ApJSS 73, 685 
\bibitem[1977]{Wal77}
Walborn, N.R., 1977, ApJ 215, 53 
\bibitem[1982]{Wal82}
Walborn, N.R., 1982, ApJ 256, 452 
\bibitem[1990]{Wal90}
Walborn, N.R., Fitzpatrick, E.L., 1990, PASP 102, 379 
\bibitem[1995]{Walk}
Walker, A.R., 1995, AJ 110, 638
\bibitem[1990]{Wes90}
Westerlund, B.E., 1990, A\&AR 2, 29 
\bibitem[1987]{Wol87}
Wolf, B., Stahl, O., Seifert, W., 1987, A\&A 186, 182 

\end{thebibliography}
\end{document}